\documentclass[11pt,a4paper]{article}
\pdfoutput=1 

\usepackage[table]{xcolor}
\usepackage{colortbl}
\RequirePackage{ifpdf} 
\usepackage{amsmath} 
\usepackage{mathtools}

\usepackage{jheppub}
\usepackage{pstricks}
\usepackage[final]{pdfpages} 
\usepackage{ifpdf} 
\usepackage{slashed}

\usepackage[normalem]{ulem}
\usepackage{color}
\usepackage{xcolor}
\definecolor{urlblue}{rgb}{0.2,0.4,0.7}
\definecolor{citegreen}{rgb}{0,0.6,0.2}
\definecolor{linkred}{rgb}{0.9,0.2,0.1}
\usepackage{hyperref}
\hypersetup{
colorlinks=true, citecolor=citegreen, linkcolor=blue, urlcolor=urlblue}

\usepackage{graphics}
\usepackage{etoolbox} 
\usepackage{fixmath}
\usepackage{psfrag}

\usepackage{notoccite} 

\usepackage{amsfonts}
\usepackage{autobreak}
\usepackage{marginnote}
\usepackage{enumitem}
\usepackage{appendix}
\usepackage{caption} 
\newcommand{\NOdisplay}[1]{ }

\def\MSbar{\overline{\mathrm{MS}}}
\def\TR{{\displaystyle \mathrm{T}_{F}}}
\def\gJJ{\gamma_{{\scriptscriptstyle J}}}
\def\gFJ{\gamma_{{\scriptscriptstyle FJ}}}
\def\gFF{\gamma_{{\scriptscriptstyle F\tilde{F}}}}

\def\qgraf{{\fontfamily{qcr}\selectfont
QGRAF}}
\def\form{{\fontfamily{qcr}\selectfont
FORM}}
\def\fermat{{\fontfamily{qcr}\selectfont
fermat}}
\def\diagen{{\fontfamily{qcr}\selectfont
DiaGen}}
\def\idsolver{{\fontfamily{qcr}\selectfont
IdSolver}}
\def\forcer{{\fontfamily{qcr}\selectfont
Forcer}}
\def\reduze{{\fontfamily{qcr}\selectfont
REDUZE2}}

\def\litered{{\fontfamily{qcr}\selectfont
LiteRed}}
\def\fire{{\fontfamily{qcr}\selectfont
FIRE6}}

\def\mathematica{{\fontfamily{qcr}\selectfont
Mathematica}}

\title{Renormalization of the flavor-singlet axial-vector current and its anomaly in dimensional regularization}

\author{Taushif Ahmed$^{a}$, Long Chen$^{b}$ and Micha\l{} Czakon$^{b}$}
\emailAdd{taushif@mpp.mpg.de, longchen@physik.rwth-aachen.de, mczakon@physik.rwth-aachen.de}
\affiliation{
$^a$Dipartimento di Fisica and Arnold-Regge Center, Universit\`a di Torino, 
\\ and INFN, Sezione di Torino, Via Pietro Giuria 1, I-10125 Torino, Italy\\
$^b$Institut f\"ur Theoretische Teilchenphysik und Kosmologie, RWTH Aachen University,\\ Sommerfeldstr.~16, 52056 Aachen, Germany}

\preprint{TTK-21-04,~P3H-21-004}

\abstract{
The renormalization constant $Z_J$ of the flavor-singlet axial-vector current with a non-anticommuting $\gamma_5$ in dimensional regularization is determined to order $\alpha_s^3$ in QCD with massless quarks. 
The result is obtained by computing the matrix elements of the operators appearing in the axial-anomaly equation $\big[\partial_{\mu} J^{\mu}_{5} \big]_{R} = \frac{\alpha_s}{4 \pi}\, n_f\, \TR \,  \big[F \tilde{F} \big]_{R}$ between the vacuum and a state of two (off-shell) gluons to 4-loop order. 
Furthermore, through this computation, the equality between the $\MSbar$ renormalization constant $Z_{F\tilde{F}}$ associated with the operator $\big[F \tilde{F} \big]_{R}$ and that of $\alpha_s$ is verified explicitly to hold true at 4-loop order. 
This equality automatically ensures a relation between the respective anomalous dimensions, $\gJJ = \frac{\alpha_s}{4 \pi}\, n_f\, \TR \, \gFJ $, at order $\alpha_s^4$ given the validity of the axial-anomaly equation which was used to determine the non-$\MSbar$ piece of $Z_J$ for the particular $\gamma_5$ prescription in use.
}

\begin{document}
\allowdisplaybreaks[4]
\unitlength1cm
\keywords{}
\maketitle
\flushbottom

\section{Introduction}
\label{sec:intro}

Dimensional regularization (DR)~\cite{tHooft:1972tcz,Bollini:1972ui} has become the choice of regularization framework that underlies most of the modern higher order perturbative calculations in Quantum Chromodynamics (QCD).
This is due to its many celebrated features. 
For example, it manifestly preserves the gauge symmetry and Lorentz invariance of QCD, renders all loop integrals invariant under arbitrary loop momentum shifts, and allows one to handle both ultraviolet (UV) and infrared/collinear (IR) divergences in the same manner without introducing separate mass scales.

On the other hand, ever since the introduction of DR, it has been recognized that special attention is required in the treatment of  $\gamma_5$, an  intrinsically 4-dimensional object.
At the root of the issue is that a fully anticommuting $\gamma_5$ is algebraically incompatible with the Dirac algebra in D ($\neq 4$) dimensions.
However, the anticommutativity of $\gamma_5$ is essential for the concept of chirality of spinors in 4 dimensions and (non-anomalous) chiral symmetries in quantum field theory (QFT).
Furthermore, there is one well-known subtlety related to $\gamma_5$ and chiral symmetries in QFT, namely, that the flavor-singlet axial-vector current\footnote{Below we denote (flavor-singlet) axial-vector current simply by ``(singlet) axial current'' for brevity.}, defined with $\gamma_5$, exhibits a quantum anomaly, the axial or Adler-Bell-Jackiw (ABJ) anomaly~\cite{Adler:1969gk,Bell:1969ts}, in its divergence.
An anticommuting $\gamma_5$ together with the invariance of loop integrals under arbitrary loop-momentum shifts, however, leads to the absence of the axial anomaly, which can be easily checked at one-loop order which subsequently holds to all orders, due to the Adler-Bardeen theorem~\cite{Adler:1969er}.

A proper and practical treatment of $\gamma_5$ in DR must reproduce the known physical properties of $\gamma_5$-related objects, established in a more traditional or physical regularization scheme (e.g.~the Pauli-Villars regularization~\cite{Pauli:1949zm}).
To be more specific, the following criteria characterise a consistent $\gamma_5$ prescription in dimensional regularization: 
\begin{enumerate}[label=(\roman*)]
\item 
in the case of axial-anomaly-free Feynman diagrams (e.g.~with all fermion lines having an even number of axial currents), the anticommutativity of $\gamma_5$ must be effectively ensured if not directly asserted from the outset; 
\item 
in the case of Feynman diagrams with a genuine axial anomaly, the all-order axial-anomaly equation~\cite{Adler:1969gk,Adler:1969er} must be preserved. 
\end{enumerate}

Despite the aforementioned difficulties, various $\gamma_5$ prescriptions in DR have been developed in the literature~\cite{tHooft:1972tcz,Akyeampong:1973xi,Breitenlohner:1977hr,Bardeen:1972vi,Chanowitz:1979zu,Gottlieb:1979ix,Ovrut:1981ne,Espriu:1982bw,Buras:1989xd,Kreimer:1989ke,Korner:1991sx,Larin:1991tj,Larin:1993tq,Jegerlehner:2000dz,Moch:2015usa,Zerf:2019ynn} over a span of nearly 50 years, albeit each with its own pros and cons.
In spite of the amount of work, the subject is still capable of holding some amusing surprises, as recently demonstrated in ref.~\cite{Ahmed:2020kme}, where a set of individually finite Feynman diagrams with an axial current regularized with a non-anticommuting $\gamma_5$ in DR, derived from the heavy-top effective Lagrangian for $q\bar{q} \rightarrow ZH$, calls for additional manual corrections.~\footnote{Applications of a non-anticommuting $\gamma_5$ to the radiative corrections in electro-weak theory or even theories with beyond-Standard-Model interactions are more cumbersome, see e.g.~\cite{Barroso:1990ti,Jegerlehner:2000dz,BeluscaMaito:2020ala}.} 
~\\

In this article, we use a particular variant of the original $\gamma_5$ prescription (and $\gamma^{\mu}\gamma_5$)~\cite{tHooft:1972tcz,Akyeampong:1973xi,Breitenlohner:1977hr} in DR, as constructively defined in  refs.~\cite{Larin:1991tj,Larin:1993tq}, where $\gamma_5$ does not anticommute with all $\gamma^{\mu}$ in $D$ dimensions. 
We compute the renormalization constants of the local composite operators in the axial-anomaly equation with this $\gamma_5$ prescription within QCD \footnote{We note that different renormalization constants will arise depending on how exactly the formal $\gamma_5$ is defined in $D$ dimensions, notably for anticommuting and non-anticommuting $\gamma_5$ prescriptions and, more subtly, among different variants of dimensional regularization with a non-anticommuting $\gamma_5$; see e.g.~\cite{Gnendiger:2017rfh}.}.
Closely related to the axial anomaly, it was known already since the early work~\cite{Adler:1969gk} that these operators, namely, the divergence of the axial current operator and the axial-anomaly operator, require additional non-trivial renormalizations in an Abelian gauge theory even with Pauli-Villars regularization (where $\gamma_5$ is just the standard 4-dimensional anticommuting Dirac matrix). 
The situation becomes only more involved with a non-anticommuting $\gamma_5$ prescription in DR due to the appearance of spurious terms originating from the apparent loss of $\gamma_5$'s anticommutativity.
With a constructive non-anticommuting $\gamma_5$ in DR, every Feynman diagram derived from the Feynman rules of the theory under consideration is mathematically unambiguously determined.
However, the price one has to pay in exchange for this mathematical clarity, besides the increment in the cost of computing traces with a large number of $\gamma$ matrices, is that one needs a non-trivial renormalization to eliminate some spurious terms whose presence spoils the correct form of chiral Ward identities, both for the anomalous and non-anomalous axial currents. 
For this reason, a pure $\MSbar$ renormalization~\cite{Bardeen:1978yd} is typically insufficient for $\gamma_5$-related objects in QFT~\cite{Chanowitz:1979zu,Trueman:1979en,Gottlieb:1979ix,Espriu:1982bw,Larin:1991tj,Larin:1993tq,Bos:1992nd}.

As far as the work of this article is concerned, the renormalization constants for a flavor non-singlet axial current and a pseudoscalar current were determined in DR to $\mathcal{O}(\alpha_s^3)$ in refs.~\cite{Larin:1991tj,Larin:1993tq}, whereas that of the flavor-singlet axial current was computed only to $\mathcal{O}(\alpha_s^2)$. 
In addition, these renormalization constants including the $\epsilon$-suppressed terms at $\mathcal{O}(\alpha_s^2)$ were determined and discussed in the calculation of polarized splitting functions~\cite{Matiounine:1998re,Vogt:2008yw,Moch:2014sna,Behring:2019tus}.
The renormalization constants needed for the axial-anomaly operator were known to $\mathcal{O}(\alpha_s^2)$ from ref.\cite{Bos:1992nd,Larin:1993tq} and were computed to $\mathcal{O}(\alpha_s^3)$ in refs.~\cite{Zoller:2013ixa,Ahmed:2015qpa}.
In this article, we extend the knowledge of the renormalization constant of the singlet axial current to $\mathcal{O}(\alpha_s^3)$ by computing the matrix elements of the operators in the axial-anomaly equation between the vacuum and a pair of (off-shell) gluons to 4-loop order. 
With the same computational set-up, the $\MSbar$ renormalization constant associated with the axial-anomaly operator is determined to $\mathcal{O}(\alpha_s^4)$ and subsequently shown to coincide with that of $\alpha_s$, a result claimed to hold in ref.~\cite{Breitenlohner:1983pi} using the background field method~\cite{Abbott:1980hw}.
The consequence of this equality in QCD in the low-energy region was discussed to $\mathcal{O}(a^2_s)$ in ref.~\cite{Chetyrkin:1998mw} in the context of a heavy-top effective Lagrangian describing the interactions of a CP-odd scalar with gluons.
~\\

The article is organized as follows.
In section~\ref{sec:aaop}, we specify the prescription for $\gamma_5$ in general and the (singlet) axial current in particular. 
Furthermore, we describe the renormalization of operators in the axial-anomaly equation.
In section~\ref{sec:proj}, we explain in detail the simple projector employed to project out the matrix elements of operators involved with the chosen special kinematics needed to extract the relevant renormalization constants. 
The computational details are exposed in section~\ref{sec:compdet}, including the renormalization procedure, followed by our results on the renormalization constants of the singlet axial-current, the axial-anomaly operator, with a discussion of their implications in section~\ref{sec:res}. 
We conclude in section~\ref{sec:conc}.

\section{Renormalization of the singlet axial current}
\label{sec:aaop}

Multi-loop calculations in QFT involving axial currents in dimensional regularization~\cite{tHooft:1972tcz,Bollini:1972ui} face the issue of defining  inherently 4-dimensional objects, e.g.~Dirac's $\gamma_5$ (and the closely related Levi-Civita tensor $\epsilon^{\mu\nu\rho\sigma}$), properly in $D$ dimensions. 
In this article, we use the non-anticommuting definition of $\gamma_5$ in dimensional regularization, originally introduced by 't Hooft-Veltman~\cite{tHooft:1972tcz} and Breitenlohner-Maison~\cite{Breitenlohner:1977hr}
\begin{align}
\label{eq:gamma5}
	\gamma_5=-\frac{i}{4!}\epsilon^{\mu\nu\rho\sigma}\gamma_{\mu}\gamma_{\nu}\gamma_{\rho}\gamma_{\sigma}\,,
\end{align}
but with the Levi-Civita tensor\footnote{We use the convention $\epsilon^{0123} = -\epsilon_{0123} = +1$.} $\epsilon^{\mu\nu\rho\sigma}$ treated according to refs.~\cite{Larin:1991tj,Zijlstra:1992kj,Larin:1993tq}.
In particular, the contraction of the $\epsilon^{\mu\nu\rho\sigma}$ tensor in eq.(\ref{eq:gamma5}) with the one from an external projector will be done according to the usual mathematical identity in 4 dimensions but with the Lorentz indices of the resulting spacetime metric tensors all taken as $D$-dimensional.
Due to this different treatment of $\epsilon^{\mu\nu\rho\sigma}$ in the above constructive definition of $\gamma_5$ in DR (compared to the original), this prescription is sometimes called Larin's prescription in the literature.   
The spacetime dimension $D$ is parameterized conveniently as $D \equiv 4 - 2 \epsilon$.

In order to define the hermitian axial current correctly with a non-anticommuting $\gamma_5$ one must ``symmetrize'' the matrix product, $\gamma^{\mu}\gamma_5 \rightarrow \frac{1}{2} \Big(\gamma^{\mu}\gamma_5-\gamma_5\gamma^{\mu}\Big)$.
With eq.~\eqref{eq:gamma5}, one obtains~\cite{Akyeampong:1973xi}
\begin{align}
\label{eq:gamma5-axial}
	\gamma^{\mu}\gamma_5 = -\frac{i}{3!} \epsilon^{\mu\nu\rho\sigma} \gamma_{\nu} \gamma_{\rho} \gamma_{\sigma}.
\end{align}
The $\gamma_5$ defined by eq.~\eqref{eq:gamma5} no longer anticommutes with all $\gamma^{\mu}$ in $D$ dimensions, which has profound consequences in higher-order computations involving axial currents in DR. 
The apparent violation of the anticommutativity of $\gamma_5$ with $\gamma^{\mu}$ in the ``evanescent'' $4-D=2\epsilon$ dimensional space gives rise to some additional spurious UV-divergence-originated terms in dimensionally regularized amplitudes, which typically leads to the loss of the correct form of chiral Ward identities for both non-singlet and singlet axial currents.
This calls for non-trivial UV renormalization of $\gamma_5$-related objects \cite{Chanowitz:1979zu,Trueman:1979en,Kodaira:1979pa,Espriu:1982bw,Collins:1984xc,Larin:1991tj,Larin:1993tq,Bos:1992nd}.
Due to the aforementioned technical origin of these spurious terms, the amendment is typically realized in the form of some additional UV renormalization constants on top of the pure $\MSbar$ renormalization constants, introduced specifically for the axial or $\gamma_5$-dependent part of the amplitudes.
~\\

\begin{figure}[htbp]
\begin{center}
\includegraphics[scale=0.85]{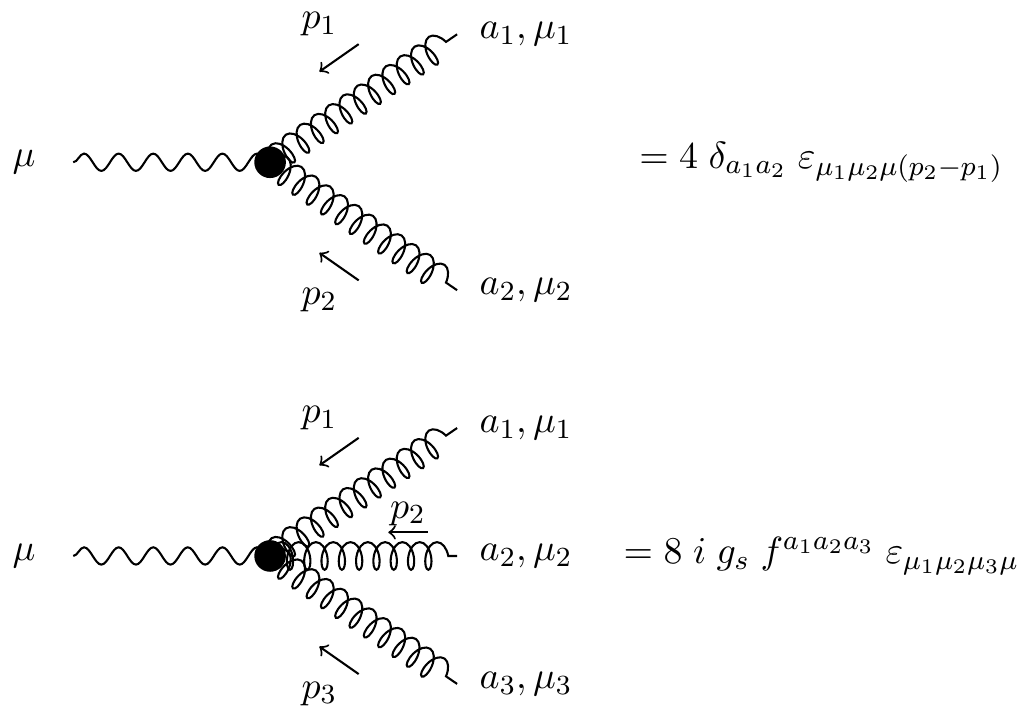}
\caption{Feynman rules for the current $K_\mu$ in eq.~\eqref{eq:Kcurrent}.}
\label{fig:feynman_rhs}
\end{center}
\end{figure}

In this article, we work in QCD with $n_f$ massless quarks.
As nicely summarized in refs.~\cite{Larin:1991tj,Larin:1993tq}, the properly renormalized singlet axial current, with $\gamma^{\mu}\gamma_5$ defined in eq.~\eqref{eq:gamma5-axial}, can be written as 
\begin{eqnarray} 
\label{eq:J5uvr}
\left[ J^{\mu}_{5}\right]_{R} &=& \sum_{\psi_B} Z_J \, \mu^{4-D}\, \bar{\psi}_{B}  \, \gamma^{\mu}\gamma_5 \, \psi_{B} \nonumber\\
&=& \sum_{\psi_B} Z^{f}_{5} \, Z^{ms}_{5} \, \mu^{4-D}\, \bar{\psi}_{B}  \, \frac{-i}{3!} \epsilon^{\mu\nu\rho\sigma} \gamma_{\nu} \gamma_{\rho} \gamma_{\sigma} \, \psi_{B} \,,
\end{eqnarray}
where $\psi_{B}$ denotes a bare quark field with mass dimension $(D-1)/2$ and the subscript $R$ at a square bracket denotes operator renormalization. 
A factor $\mu^{4-D}$ in the mass scale $\mu$ of dimensional regularization has been inserted in eq.~(\ref{eq:J5uvr}) in order for the mass dimension of the r.h.s.\ operator be equal to the canonical dimension of the l.h.s.\ in four dimensions. 
The sum extends over all $n_f$ quark fields. 
Here and below $J^{\mu}_{5}$ denotes the bare flavor-singlet axial current.
It is known to renormalize multiplicatively~\cite{Adler:1969gk,Trueman:1979en}, as it is the only local composite current operator in the context of QCD that has the correct mass dimension and conserved quantum numbers (which are preserved under renormalization). 
The factor $Z_{J} \equiv Z^{f}_{5} \, Z^{ms}_{5}$ denotes the UV renormalization constant of the current,
conveniently parameterized as the product of a pure $\MSbar$-renormalization part $Z^{ms}_{5}$ and an additional finite renormalization 
 factor $Z^{f}_{5}$.
The latter is needed to restore the correct form of the axial Ward identity, namely, the all-order axial-anomaly equation~\cite{Adler:1969gk,Adler:1969er}, which reads in terms of renormalized local composite operators\footnote{When inserted into a Green's function, the time component of the derivative generates \textit{contact} terms in the respective Ward identity.}
\begin{eqnarray} 
\label{eq:ABJanomalyEQ}
\big[\partial_{\mu} J^{\mu}_{5} \big]_{R} = a_s\, n_f\, \TR \,  \big[F \tilde{F} \big]_{R}\,,
\end{eqnarray}
where $\TR=1/2$, $F \tilde{F} \equiv  - \epsilon^{\mu\nu\rho\sigma} F^a_{\mu\nu} F^a_{\rho\sigma} = \epsilon_{\mu\nu\rho\sigma} F^a_{\mu\nu} F^a_{\rho\sigma}$  denotes the contraction of the field strength tensor $F^a_{\mu\nu} = \partial_{\mu} A_{\nu}^{a} - \partial_{\nu} A_{\mu}^{a} + g_s \,  f^{abc} A_{\mu}^{b} A_{\nu}^{c}$ of the gluon field $A_\mu^a$ with its \textit{dual} form.
We use the shorthand notation $a_s \equiv \frac{\alpha_s}{4 \pi} = \frac{g_s^2}{16 \pi^2}$ for the QCD coupling, and $f^{abc}$ denotes the structure constants of the non-Abelian color group of QCD. 
In contrast to the l.h.s. of \eqref{eq:ABJanomalyEQ}, the renormalization of the axial-anomaly operator $F \tilde{F}$ is not strictly multiplicative (as known from ref.~\cite{Adler:1969gk}), but involves mixing with the divergence of the axial current operator~\cite{Espriu:1982bw,Breitenlohner:1983pi}, 
\begin{eqnarray} 
\label{eq:FFuvr}
\big[F \tilde{F} \big]_{R} = Z_{F\tilde{F}} \, \mu^{4-D}\, \big[F \tilde{F} \big]_{B} \,+\, 
 Z_{FJ} \, \mu^{4-D}\, \big[\partial_{\mu} J^{\mu}_{5} \big]_{B} \,,
\end{eqnarray}
where the subscript $B$ implies that the fields in the local composite operators are bare, and a mass-dimension compensation factor $\mu^{4-D}$ has been introduced as in eq.~(\ref{eq:J5uvr}).
In the computation of the matrix elements of the axial-anomaly operator $F \tilde{F}$, we employ its equivalent form in terms of the divergence of the ``axial gluon current'' $K^{\mu}$ (a Chern-Simons 3-form), namely 
\begin{eqnarray} 
\label{eq:Kcurrent}
F \tilde{F}  &=& \partial_{\mu} K^{\mu} \nonumber\\
 &=& \partial_{\mu} \left(-4 \,\epsilon^{\mu\nu\rho\sigma} \,\left(A_{\nu}^{a} \partial_{\rho} A_{\sigma}^{a} \,+\, g_s\,\frac{1}{3} f^{abc} A_{\nu}^{a} A_{\rho}^{b} A_{\sigma}^{c} \right) \right)\,,
\end{eqnarray}
where, unlike $J^{\mu}_{5}$, the current $K^{\mu} \equiv -4 \, \epsilon^{\mu\nu\rho\sigma} \,\left(A_{\nu}^{a} \partial_{\rho} A_{\sigma}^{a} \,+\, g_s\,\frac{1}{3} f^{abc} A_{\nu}^{a} A_{\rho}^{b} A_{\sigma}^{c} \right)$ is not a gauge-invariant object~\cite{Espriu:1982bw,Breitenlohner:1983pi}.
The practical advantage of using this alternative form of the axial-anomaly operator in perturbative computations is already clear from the fact that the number of gluon fields therein is one less than in its original defining form.
We derive the Feynman rules for the r.h.s. of eq.~\eqref{eq:ABJanomalyEQ} directly based on eq.~\eqref{eq:Kcurrent}. 
These rules are listed in figure~\ref{fig:feynman_rhs}, and are used in our computations to be presented in the following sections.

The renormalization of the operators $\partial_{\mu} J^{\mu}_{5}$ and $F \tilde{F}$ specified, respectively, in eq.~\eqref{eq:J5uvr} and eq.~\eqref{eq:FFuvr} can be arranged into the following matrix form 
\begin{eqnarray}
\label{eq:Zsmatrix}
\begin{pmatrix}
\big[\partial_{\mu} J^{\mu}_{5} \big]_{R}\\
\big[F \tilde{F}\big]_{R}
\end{pmatrix}
= \mu^{4-D}\,
\begin{pmatrix}
Z_{J} &  0 \\
Z_{FJ} &  Z_{F\tilde{F}}
\end{pmatrix}
\cdot 
\begin{pmatrix}
\big[\partial_{\mu} J^{\mu}_{5} \big]_{B}\\
\big[F \tilde{F}\big]_{B}
\end{pmatrix}\,.
\end{eqnarray}
The matrix of anomalous dimensions of these two renormalized operators is defined by 
\begin{eqnarray}
\label{eq:AMDmatrix}
\frac{\mathrm{d}}{\mathrm{d}\, \ln \mu^2}\,
\begin{pmatrix}
\big[\partial_{\mu} J^{\mu}_{5} \big]_{R}\\
\big[F \tilde{F}\big]_{R}
\end{pmatrix}
= 
\begin{pmatrix}
\gJJ &  0 \\
\gFJ &  \gFF
\end{pmatrix}
\cdot 
\begin{pmatrix}
\big[\partial_{\mu} J^{\mu}_{5} \big]_{R}\\
\big[F \tilde{F}\big]_{R}\,.
\end{pmatrix}
\end{eqnarray}
Inserting eq.~\eqref{eq:Zsmatrix} into eq.~\eqref{eq:AMDmatrix} leads to the following expressions for anomalous dimensions in terms of the renormalization constants:
\begin{eqnarray}
\label{eq:AMDinZs}
\begin{pmatrix}
\gJJ &  0 \\
\gFJ &  \gFF
\end{pmatrix}
&=&  
\begin{pmatrix}
\epsilon \,~\quad &  0 \\
0 \,~\quad &  \epsilon
\end{pmatrix}
+
\begin{pmatrix}
\frac{\mathrm{d}\, Z_{J}}{\mathrm{d}\, \ln \mu^2}\,  &  0 \\
\frac{\mathrm{d}\, Z_{FJ}}{\mathrm{d}\, \ln \mu^2} &  
\frac{\mathrm{d}\, Z_{F\tilde{F}}}{\mathrm{d}\, \ln \mu^2}
\end{pmatrix}
\cdot 
\begin{pmatrix}
Z_{J} &  0 \\
Z_{FJ} &  Z_{F\tilde{F}}
\end{pmatrix}^{-1}
\nonumber\\
&=& 
\begin{pmatrix}
\epsilon + \frac{\mathrm{d}\, \ln Z_{J}}{\mathrm{d}\, \ln \mu^2}\,  &  0 \\
\frac{1}{Z_J}\frac{\mathrm{d}\, Z_{FJ}}{\mathrm{d}\, \ln \mu^2}
- \frac{Z_{FJ}}{Z_J} \frac{\mathrm{d}\, \ln Z_{F\tilde{F}}}{\mathrm{d}\, \ln \mu^2} &  
\epsilon +\frac{\mathrm{d}\, \ln Z_{F\tilde{F}}}{\mathrm{d}\, \ln \mu^2}
\end{pmatrix}
\,.
\end{eqnarray}

Before we plunge into the technical details of the computation, let us quickly comment on our choice of computing the matrix element of the flavor-singlet axial current $J^{\mu}_{5}$, or rather its divergence  $\partial_{\mu} J^{\mu}_{5}$, between the vacuum and a pair of off-shell gluons. 
The factor $Z_{J} \equiv Z^{f}_{5} \, Z^{ms}_{5}$ in QCD is an overall UV-renormalization constant for the axial-current operator, which by itself carries no Lorentz index and does not distinguish the components of the axial current.  
The very same universal UV-renormalization constant for the axial-current operator, at least the pure $\MSbar$ part $Z^{ms}_{5}$, can be determined by computing the matrix elements of $J^{\mu}_{5}$, or alternatively those of its ``scalar component'' $\partial_{\mu} J^{\mu}_{5}$, between the vacuum and a pair of external quarks, or any other external states, provided these matrix elements are not vanishing with the chosen kinematics.
To further fix the additional non-$\MSbar$ finite $Z^{f}_{5}$ for an anomalous singlet-axial current, one could compute the matrix elements of both sides of the axial-anomaly equation~\eqref{eq:ABJanomalyEQ} between the same external states, and subsequently demand the matching between the l.h.s. and r.h.s. matrix elements order by order in perturbation theory.
Having in mind that, besides determining $Z_{J} \equiv Z^{f}_{5} \, Z^{ms}_{5}$, we would also like to verify the equality between the $\MSbar$ renormalization constant $Z_{F\tilde{F}}$ and that of $a_s$ at $\mathcal{O}(\alpha_s^4)$, we choose to only compute the matrix elements with two external gluons.
However, to determine $Z^f_{5}$ to $\mathcal{O}(a^N_s)$ from this choice of external states, an (N+1)-loop computation is needed, because in this case the matrix element of $\partial_{\mu} J^{\mu}_{5}$ starts at one-loop order (i.e.~the anomalous fermion-triangle diagram).
In particular, $Z^f_{5}$ was determined to $\mathcal{O}(a^2_s)$ from a 3-loop computation of these matrix elements in ref.~\cite{Larin:1993tq}, while in this article we will extend this result to one order higher in $a_s$.

\section{The axial-anomaly projector}
\label{sec:proj}

\begin{figure}[htbp]
\begin{center}
\includegraphics[scale=0.8]{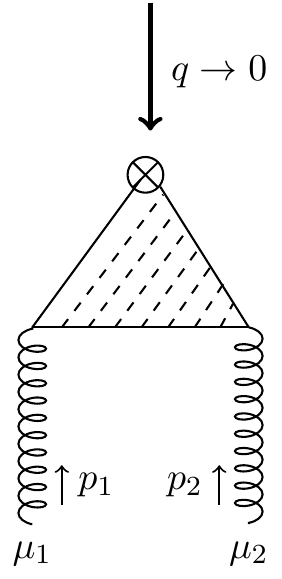}
\caption{The kinematic configuration selected for the computations of matrix elements of the operators appearing in the axial-anomaly equation.}
\label{fig:kinematics}
\end{center}
\end{figure}

We choose to determine the renormalization constant $Z_{J} \equiv Z^{f}_{5} \, Z^{ms}_{5}$ of the singlet axial current to $\mathcal{O}(a^3_s)$, and $Z_{F\tilde{F}}$ to $\mathcal{O}(a^4_s)$, in DR by computing matrix elements of operators appearing in the axial-anomaly equation between the vacuum and a pair of off-shell gluons evaluated at a specifically chosen single-scale kinematic configuration~\cite{Bos:1992nd,Larin:1993tq}, as illustrated in figure~\ref{fig:kinematics}.

Let us denote by $\langle 0| \hat{\mathrm{T}}\left[ J_{5}^{\mu}(y) \, A_a^{\mu_1}(x_1) \, A_a^{\mu_2}(x_2) \right] |0 \rangle |_{\mathrm{amp}}$ the amputated one-particle irreducible (1PI) vacuum expectation value of the time-ordered covariant product of the (singlet) axial current and two gluon fields in coordinate space with open Lorentz indices. 
Subsequently, we introduce the following rank-3 matrix element $\Gamma^{\mu \mu_1 \mu_2}_{lhs}(p_1, p_2)$ in momentum space, defined by
\begin{eqnarray}
\label{eq:Glhs1PI}
\Gamma^{\mu \mu_1 \mu_2}_{lhs}(p_1, p_2) \equiv 
\int d^4 x  d^4 y \, e^{-i p_1 \cdot x - i q \cdot y }\,  \langle 0| \hat{\mathrm{T}}\left[ J_{5}^{\mu}(y) \, A_a^{\mu_1}(x) \, A_a^{\mu_2}(0) \right] |0 \rangle |_{\mathrm{amp}} 
\end{eqnarray}
where translation invariance 
has been used to shift the coordinate of one gluon field to the origin, and the resulting total momentum conservation factor, ensuring $p_2 = -q - p_1$, is implicit. 
At leading order, this is just the matrix element corresponding to the famous one-loop fermion-triangle diagram~\cite{Adler:1969gk,Bell:1969ts} with the polarization vectors of the external gluons stripped off and without imposing on-shell constraints on the incoming gluon momenta.
The higher order corrections consist of all 1PI diagrams with amputated external gluon legs to the specific loop order in question.

However, rather than performing the projection for the axial anomaly literally as devised in eq.~(19) of ref.~\cite{Larin:1993tq} where a derivative w.r.t the momentum $q$ is taken before going to the limit $q_\mu \rightarrow 0$~(see also ref.~\cite{Bos:1992nd}), we use instead the following projector 
\begin{eqnarray}
\label{eq:anomalyprojector}
\mathcal{P}_{\mu \mu_1 \mu_2} = -\frac{1}{6 \, p_1 \cdot p_1} \, \epsilon_{\mu\mu_1\mu_2\nu}\, p_1^{\nu} \,,
\end{eqnarray}
and directly project $\Gamma^{\mu \mu_1 \mu_2}(p_1, -p_1)$ onto this structure right at $q = 0$ (i.e.~$p_2=-p_1$). 
This leads to the scalar mass-dimensionless matrix element
\begin{eqnarray}
\label{eq:smeL}
\mathcal{M}_{lhs} =  \mathcal{P}_{\mu \mu_1 \mu_2} \,  \Gamma^{\mu \mu_1 \mu_2}_{lhs}(p_1, -p_1) \,.
\end{eqnarray}
In fact, eq.(\ref{eq:anomalyprojector}) encodes the one and only ``physical'' Lorentz structure of $\Gamma^{\mu \mu_1 \mu_2}_{lhs}(p_1, -p_1) $ surviving at the chosen kinematics $q=0, p_2 = - p_1$ under the condition of having one Levi-Civita tensor and being Bose-symmetric.
Consequently, the projector $\mathcal{P}_{\mu \mu_1 \mu_2}$ projects out the form factor in front of this unique structure, which is not vanishing in the limit $q=0$.
Strictly speaking, with all Lorentz indices taken to be $D$-dimensional as stated below eq.~\eqref{eq:gamma5}, the squared norm of the Lorentz structure $\epsilon_{\mu\mu_1\mu_2\nu} p_1^{\nu}$ is equal to (6-11$D$+6$D^2$-$D^3$)$p_1\cdot p_1$, but it is known~\cite{Chen:2019wyb,Ahmed:2019udm} that the parameter $D$ here can be safely set to 4 consistently throughout the computation in DR without problem~\footnote{This has already been employed extensively in ref.~\cite{Chen:2019wyb} for devising projectors to project out amplitudes directly both in the linear polarization and helicity basis, and along a similar line of thinking it was used for conventional form factor projectors in ref.~\cite{Ahmed:2019udm}. In physical applications the number of these projections needed are equal to the number of independent helicity amplitudes in 4 dimensions.}.
In fact, it is also convenient to do so in our computational set-up, to be discussed in the next section, as the projection of $\mathcal{M}_{lhs}$ is separated from the evaluation and substitution of each individual loop integral therein.

The scalar quantity $\mathcal{M}_{lhs}$ is projected out and evaluated right at the limit of zero momentum $q = -(p_1 + p_2)$ 
flowing through the inserted operator $J_{5}^{\mu}(y)$, but with off-shell gluon momentum $p_1^2 \neq 0$.  
The reasons for choosing such a special single-scale kinematic configuration,  common in computations of QCD UV renormalization constants in $\MSbar$ (e.g.~in refs.~\cite{Larin:1993tp,vanRitbergen:1997va,Chetyrkin:1999pq,Czakon:2004bu,Chetyrkin:2004mf,Baikov:2016tgj,Luthe:2016xec,Chetyrkin:2017bjc}), are the following:
\begin{itemize}
\item 
the possible IR divergences in $\Gamma^{\mu \mu_1 \mu_2}_{lhs}(p_1, p_2)$ are nullified and all the poles in $\epsilon$ evaluated with off-shell momentum $p_1$ are of UV origin owing to the IR-rearrangement~\cite{Vladimirov:1979zm};
\item 
all loop integrals involved are reduced to massless propagator-type integrals which are well studied and known numerically~\cite{Smirnov:2010hd} as well as analytically~\cite{Baikov:2010hf,Lee:2011jt} to 4-loop order.
\end{itemize}
Because $p_1^2\neq 0$ is the only physical scale involved in the matrix elements under consideration, we set $p_1^2 = 1$ from the outset in our computations without loss of generality. ~\\

Although, at first sight, the projector $\mathcal{P}_{\mu \mu_1 \mu_2}$ in eq.(\ref{eq:anomalyprojector}) does not seem to have anything to do with the axial anomaly, it can be shown that with the appropriate regularity condition it is indeed equivalent to the operation devised for projecting out the anomaly in eq.(19) of ref.~\cite{Larin:1993tq}; see Appendix~\ref{append:aap}. 
The very same projector $\mathcal{P}_{\mu \mu_1 \mu_2}$ is also used in extracting a scalar quantity from the matrix element of $\big[F \tilde{F} \big]_{R}$ between the vacuum and the same external (off-shell) gluon state.
In order to be able to apply eq.~(\ref{eq:anomalyprojector}), it is crucial to use the equivalent form of the axial-anomaly operator $F \tilde{F}  = \partial_{\mu} K^{\mu}$ in terms of the divergence of the current $K^{\mu}$.
To be specific, we define in analogy to eq.~\eqref{eq:Glhs1PI}:
\begin{eqnarray}
\label{eq:Grhs1PI}
\Gamma^{\mu \mu_1 \mu_2}_{rhs}(p_1, p_2) \equiv 
\int d^4 x  d^4 y \, e^{-i p_1 \cdot x - i q \cdot y }\,  \langle 0| \hat{\mathrm{T}}\left[ K^{\mu}(y) \, A_a^{\mu_1}(x) \, A_a^{\mu_2}(0) \right] |0 \rangle |_{\mathrm{amp}} \,.
\end{eqnarray}
Comments below eq.~\eqref{eq:Glhs1PI} apply here as well.
Subsequently, one contracts $\Gamma^{\mu \mu_1 \mu_2}_{rhs}(p_1, -p_1)$ with $\mathcal{P}_{\mu \mu_1 \mu_2}$, yielding a r.h.s. counterpart to eq.~\eqref{eq:smeL}:  
\begin{eqnarray}
\label{eq:smeR}
\mathcal{M}_{rhs} =  \mathcal{P}_{\mu \mu_1 \mu_2} \,  \Gamma^{\mu \mu_1 \mu_2}_{rhs}(p_1, -p_1) \,.
\end{eqnarray}
The perturbative corrections to $\mathcal{M}_{rhs}$ in terms of Feynman diagrams must be computed using the Feynman rules involving $K^\mu$ listed in figure~\ref{fig:feynman_rhs}.

\section{Details of the computation}
\label{sec:compdet}

We are now in the position to discuss the technical aspects of the computation of the scalar projections $\mathcal{M}_{lhs}$ and $\mathcal{M}_{rhs}$ to 4-loop order in DR out of which $Z_{J} \equiv Z^{f}_{5} \, Z^{ms}_{5}$ is determined to order $\alpha_s^3$, and $Z_{F\tilde{F}}$ to order $\alpha_s^4$. 

\subsection*{The computational set-up}

\begin{figure}[htbp]
\begin{center}
\includegraphics[scale=0.8]{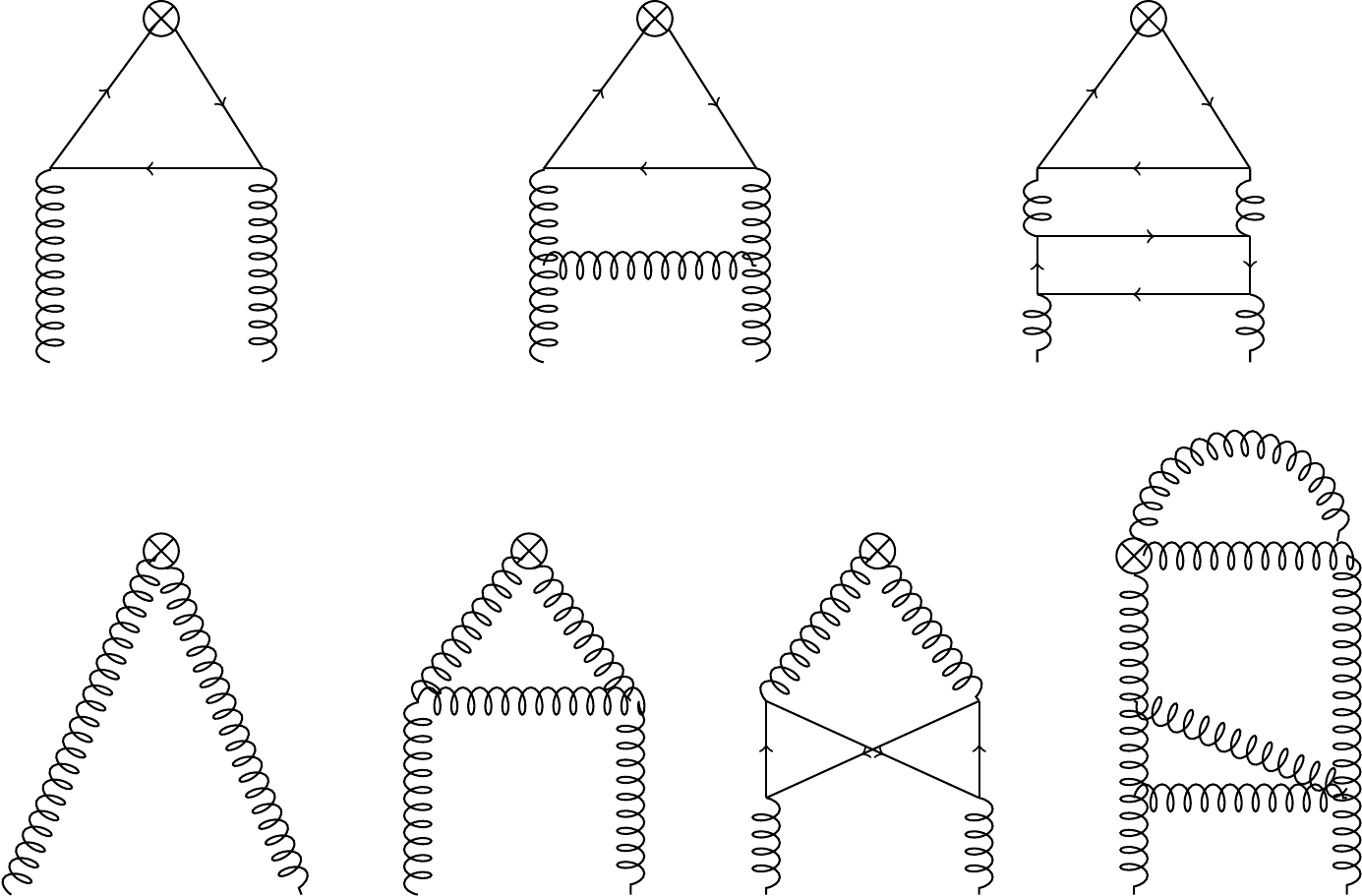}
\caption{Sample Feynman diagrams up to 3-loop order contributing to the matrix elements ${\cal M}_{lhs}$ (upper row) and ${\cal M}_{rhs}$ (lower row). 
}
\label{fig:dia123loop}
\end{center}
\end{figure}

\begin{figure}[htbp]
\begin{center}
\includegraphics[scale=0.8]{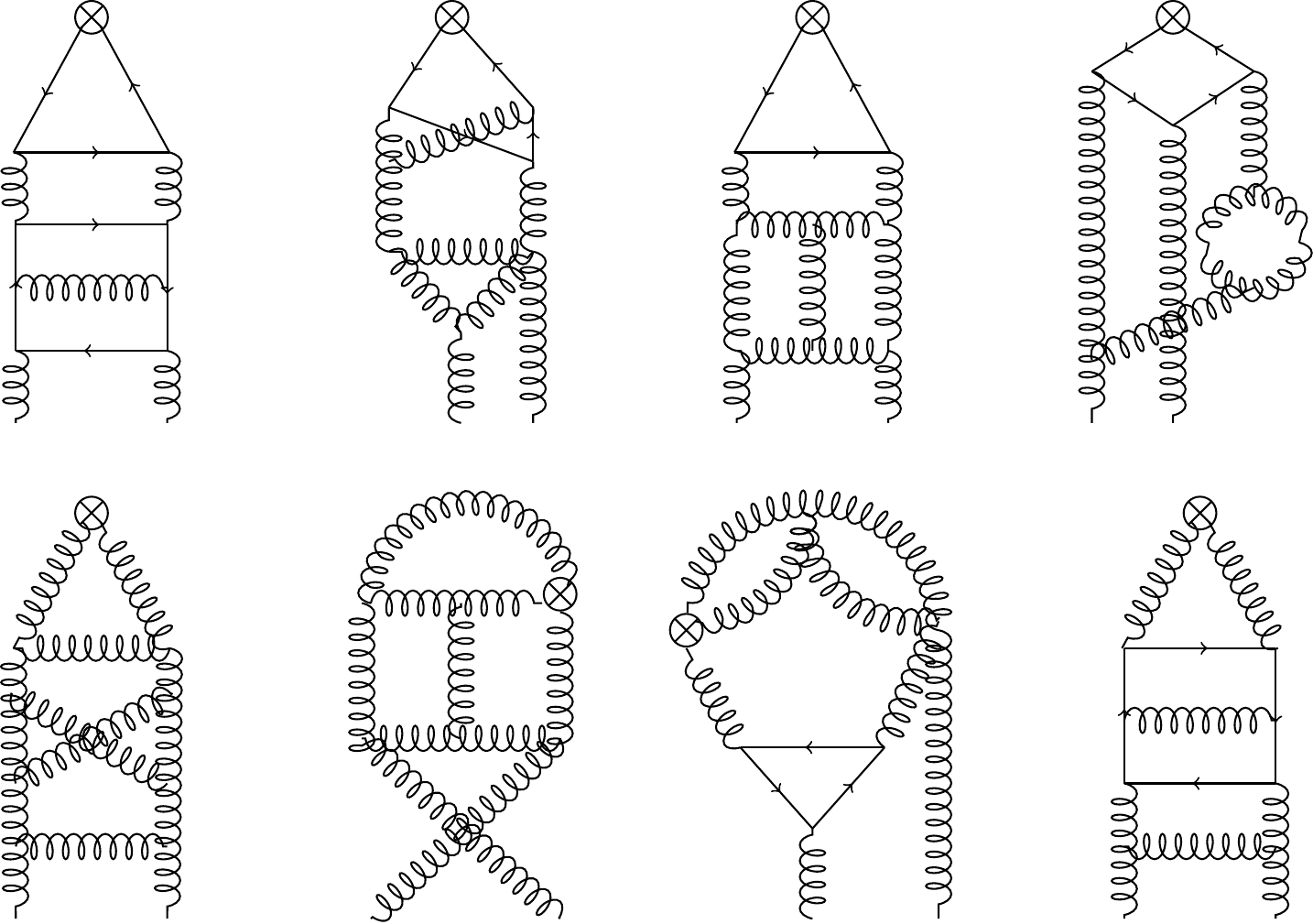}
\caption{Sample Feynman diagrams at 4-loop contributing to the matrix elements ${\cal M}_{lhs}$ (upper row) and ${\cal M}_{rhs}$ (lower row)}
\label{fig:dia4loop}
\end{center}
\end{figure}

We compute the pertubative QCD corrections to $\mathcal{M}_{lhs}$ and $\mathcal{M}_{rhs}$ in terms of Feynman diagrams, which are manipulated in the usual way.
Symbolic expressions of the contributing Feynman diagrams to 4-loop order are generated by the diagram generator \diagen~\cite{diagen}\footnote{The C++ library \diagen~provides, besides diagram generation for arbitrary Feynman rules, topological analysis tools and an interface to the C++ library \idsolver~that allows to directly apply integration-by-parts identities to the integrals occurring in the generated diagrams. \idsolver~has been originally written for the calculation of Ref.~\cite{Czakon:2004bu}, while \diagen~predates this software.}, based on the QCD Lagrangian supplemented by two additional local composite operators as discussed in section~\ref{sec:aaop}. 
The numbers of generated diagrams are listed in the second and third column of table~\ref{tab:diagN} for the l.h.s. and r.h.s. matrix elements\eqref{eq:smeL} and \eqref{eq:smeR}, respectively.
A few sample diagrams are shown in figures~\ref{fig:dia123loop}  and~\ref{fig:dia4loop}.
\begin{table}
\caption{Numbers of Feynman diagrams up to 4-loop order generated by the generators \diagen~and \qgraf. Sample diagrams to 3-loop and at 4-loop are shown in figures~\ref{fig:dia123loop} and \ref{fig:dia4loop}, respectively.}
\label{tab:diagN}
\centering
\includegraphics[width=0.75\linewidth]{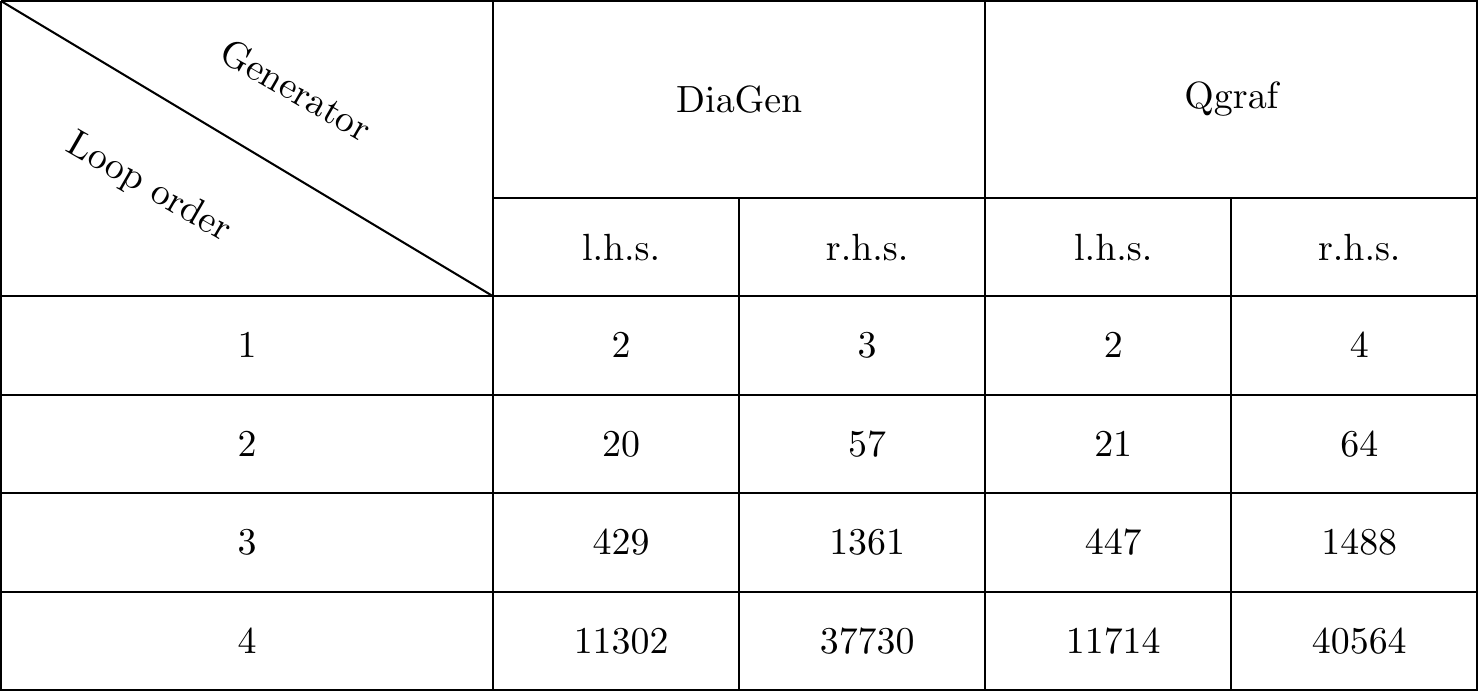}
\end{table}
Feynman rules' substitution, the color algebra with a generic SU($N_c$) group and $D$-dimensional Lorentz as well as the Dirac algebra are performed using \form~\cite{Vermaseren:2000nd}. 
The computations are done in a general Lorentz-covariant gauge for the gluon field, with the general-covariant-gauge fixing parameter $\xi$ defined through the gluon propagator $\frac{i}{k^2} \left(-g^{\mu\nu} + \xi \,\frac{k^{\mu} k^{\mu}}{k^2} \right)$, $k$ being the momentum of the gluon.
With this parametrization, the Feynman gauge corresponds to $\xi = 0$. 
As will become clear later, keeping the $\xi$ dependence in the results for $\mathcal{M}_{lhs}$ and $\mathcal{M}_{rhs}$ to at least 3-loop order, does not only serve as a check of $\xi$-independence of the renormalization constants extracted to this order, but is also necessary to achieve the full UV renormalization of $\mathcal{M}_{lhs}$ and $\mathcal{M}_{rhs}$ at the 4-loop order.

Up to 3-loop order, the reduction of loop integrals in the bare $\mathcal{M}_{lhs}$ and $\mathcal{M}_{rhs}$ to master integrals, by means of integration-by-parts identities (IBP)~\cite{Tkachov:1981wb,Chetyrkin:1981qh}, is done using \idsolver~\cite{diagen}, while analytical expressions for master integrals computed in refs.~\cite{Baikov:2010hf,Lee:2011jt} are used\footnote{In particular, we take the analytic expressions as provided in the ancillary files of ref.~\cite{Baikov:2010hf}.}. 
For the reduction of 4-loop (massless propagator-type) integrals in $\mathcal{M}_{lhs}$ and $\mathcal{M}_{rhs}$, we employ the program \forcer~\cite{Ruijl:2017cxj} which is specifically designed for the parametric reduction of this type of integrals with high performance.
To this end, we first extract the list of unreduced loop integrals appearing in our unreduced $\mathcal{M}_{lhs}$ and $\mathcal{M}_{rhs}$, the numbers of which are about $6 \times 10^4$ and $9 \times 10^4$ respectively.
We then prepare for each loop integral the \forcer~input form (which is essentially a compact notation encoding the incidence matrix of the integral's graph representation\footnote{\forcer~does not accept graphs with vertices with degree higher than 4, which are typically associated with unreduced 4-loop integrals of sub-sectors resulting from canceling certain loop propagators between denominator and numerator. 
We overcome this by introducing auxiliary graphs with only degree-3 and -4 vertices where the contracted graphs representing the sub-sector integrals can be embedded.}), and subsequently make a parallel run of \forcer~to obtain a table of generally usable IBP rules, where explicit analytical expressions of master integrals from refs.~\cite{Baikov:2010hf,Lee:2011jt} have been inserted. 
Although this may not be the most efficient way of utilizing \forcer, it does suffice for our purpose.
In fact, for the problem at hand, the process of projecting out the bare $\mathcal{M}_{lhs}$ and $\mathcal{M}_{rhs}$ at 4-loop order takes much more time than that of obtaining the IBP table (using \forcer).
For example, projecting out the bare $\mathcal{M}_{lhs}$ takes about 3 days on a computer with 24 cores (Intel\textsuperscript{\textregistered} Xeon\textsuperscript{\textregistered} Silver 4116), while with \forcer~it requires about 12 hours on a computer with 8 cores (Intel\textsuperscript{\textregistered} Xeon\textsuperscript{\textregistered} E3-1275 V2) to obtain the IBP table with analytical results.
The time needed for computing $\mathcal{M}_{rhs}$ is roughly twice as long as for computing $\mathcal{M}_{lhs}$.
Insertion of the IBP table and subsequent simplification of the final analytical results for $\mathcal{M}_{lhs}$ and $\mathcal{M}_{rhs}$ are performed using \mathematica~and \fermat~\cite{fermat} on a laptop.

In a parallel set-up serving as a cross-check, the contributing QCD diagrams are generated symbolically using the diagram generator \qgraf~\cite{Nogueira:1991ex}, with the number of diagrams listed in the fourth and fifth column in table~\ref{tab:diagN}.
The difference in the number of diagrams explicitly taken into account between the two set-ups is mainly due to the diagrams with sub loop-graphs rendered scaleless because of $q=0$, which are not explicitly excluded in the set-up with \qgraf.
The diagrams are subsequently processed through a series of in-house routines based on ~\form~in order to apply the Feynman rules, perform SU$(N_c)$ color, $D$-dimensional spinor and Lorentz algebras. 
With the projector of eq.~\eqref{eq:anomalyprojector}, $\mathcal{M}_{lhs}$ (and $\mathcal{M}_{rhs}$) is expressed as a linear combination of a large number of scalar Feynman integrals belonging to the family of massless propagator-type loop integrals at each loop order, and \reduze~\cite{Studerus:2009ye,vonManteuffel:2012np} is used to find the appropriate loop momentum shifts. 
The scalar integrals present in the projected matrix elements are reduced to linear combinations of 28 master integrals using \fire~\cite{Smirnov:2014hma,Smirnov:2019qkx} combined with \litered~\cite{Lee:2012cn,Lee:2013mka}, and the analytical results of these master integrals are taken from ref.~\cite{Baikov:2010hf}.
As expected, the time for performing the IBP reductions using the Laporta-algorithm~\cite{Laporta:2001dd} based tools is significantly longer than that needed by \forcer.
Insertion of the IBP table (in terms of symbolic masters) into $\mathcal{M}_{lhs}$ and $\mathcal{M}_{rhs}$, and subsequent simplification of the final analytical results, are performed using \mathematica~and \fermat.

Despite the superficial difference in the numbers of Feynman diagrams generated between these two set-ups, as well as the difference in the implemented technical treatments, the analytical results for the bare $\mathcal{M}_{lhs}$ and $\mathcal{M}_{rhs}$ were found to be identical, to the order needed for determining the renormalization constants presented below.

\subsection*{UV Renormalization}

In order to renormalize the bare expressions of $\mathcal{M}_{lhs}$ and $\mathcal{M}_{rhs}$ within our computational set-ups as described above, we make use of the multiplicative renormalizability of the QCD Lagrangian and of the additional local composite operators (apart from the operator mixing discussed in section~\ref{sec:aaop}).
To be definite about the notations and conventions in use, let us start with the renormalization of the bare QCD coupling ${\hat a}_s$,  
\begin{eqnarray}
\label{eq:asuvr}
{\hat a}_s \, \,S_{\epsilon} = Z_{a_s}(\mu^2)\, a_s(\mu^2) \, \mu^{2\epsilon}\,,
\end{eqnarray}
where the dimensionless renormalized coupling $a_s \equiv \frac{\alpha_s}{4 \pi} = \frac{g_s^2}{16 \pi^2}$ was already introduced below eq.~\eqref{eq:ABJanomalyEQ}.
The bare coupling ${\hat a}_s$ has mass-dimension $2\epsilon$ as is exhibited on the r.h.s. of \eqref{eq:asuvr}.
In the $\MSbar$ scheme for dimensionally-regularized loop integrals, $S_{\epsilon} = \left(4 \pi \right)^{\epsilon} e^{-\epsilon \gamma_E}$ (with $\gamma_E$ the Euler constant).
All UV poles on the r.h.s.~of eq.~\eqref{eq:asuvr} are explicitly encoded in $Z_{a_s}$ whose dependence on the scale $\mu$ is implicit and enters solely through the renormalized coupling $a_s$. (The dependence on $\mu$ of these quantities is suppressed from here onward whenever there is no confusion.)
The independence of ${\hat a}_s$ on $\mu$ implies the renormalization-group (RG) equation of $a_s$ in $D$ dimensions:
\begin{eqnarray}
\label{eq:beta}
\mu^2\frac{\mathrm{d} \ln a_s}{\mathrm{d} \mu^2} = -\epsilon - \mu^2\frac{\mathrm{d} \ln Z_{a_s}}{\mathrm{d} \mu^2} \equiv -\epsilon + \beta \,,
\end{eqnarray}
where $\beta \equiv - \mu^2\frac{\mathrm{d} \ln Z_{a_s}}{\mathrm{d} \mu^2}$ denotes the QCD beta function, i.e.,~the anomalous dimension of the renormalized $a_s$ in 4 dimensions.
Below we set $\mu = 1$ and eq.~\eqref{eq:asuvr} reduces to ${\hat a}_s = Z_{a_s}\, a_s$.
All loop integrals are evaluated in the $\MSbar$ scheme.

The multiplicative renormalizability of the QCD Lagrangian and of the axial current operator $J_5^{\mu}$ in eq.~\eqref{eq:J5uvr} implies that $\mathcal{M}_{lhs}$ can be renormalized as 
\begin{eqnarray}
\label{eq:LHSuvr}
\mathcal{M}_{lhs} &=& Z_{J} \, Z_3 \, \hat{\mathcal{M}}_{lhs}\left( {\hat a}_s\,,\, \hat{\xi} \right) \nonumber\\
&=& Z^{f}_{5} \, Z^{ms}_{5} \, Z_3 \, \hat{\mathcal{M}}_{lhs}\left( Z_{a_s}\, a_s\,,\, 1 - Z_3 + Z_3\, \xi  \right) \nonumber\\
&\equiv& Z^{f}_{5} \, \bar{\mathcal{M}}_{lhs}\,,
\end{eqnarray}
where we have suppressed the kinematic dependence of $\mathcal{M}_{lhs}$ on $p_1 \cdot p_1$ (set to 1), and as in eq.~\eqref{eq:asuvr} we used symbols with a hat to denote the unrenormalized quantities while the hat is omitted in the renormalized counterparts.
Furthermore, in the last line of the above equation, we have introduced a shorthand notation $\bar{\mathcal{M}}_{lhs} \equiv Z^{ms}_{5} \, Z_3 \, \hat{\mathcal{M}}_{lhs}\left( Z_{a_s}\, a_s\,,\, 1 - Z_3 + Z_3\, \xi  \right)$ denoting the purely $\MSbar$ renormalized form of the matrix element, which, after an additional finite renormalization by $Z^{f}_{5} $, gives the final properly renormalized $\mathcal{M}_{lhs}$.
There are two more points in the formula eq.~\eqref{eq:LHSuvr} that require comments. 
First, since the external gluons are set off-shell, we need to include the corresponding $\MSbar$ wavefunction renormalization constant $Z_3$. 
Second, the matrix element $\mathcal{M}_{lhs}$ projected out using the projector eq.~\eqref{eq:anomalyprojector} at the chosen kinematics is actually dependent on the covariant-gauge fixing parameter $\xi$, which by itself requires renormalization in QCD.
With $\xi$ defined through the gluon propagator $\frac{i}{k^2} \left(-g^{\mu\nu} + \xi \,\frac{k^{\mu} k^{\mu}}{k^2} \right)$ (in order to have less terms when inserted into diagrams with many gluons), its renormalization reads $\hat{\xi} = 1 - Z_3 + Z_3\, \xi $ where one has used the fact that $1-\hat{\xi}$ renormalizes multiplicatively with the renormalization constant $Z_{\xi}=Z_3$ (see, e.g.~\cite{Czakon:2004bu,Chetyrkin:2004mf}). 
This is why keeping the explicit $\xi$-dependence in $\hat{\mathcal{M}}_{lhs}$, at least to 3-loop order, is necessary to achieve its complete UV renormalization at 4-loop order.
After inserting the explicit expressions of the needed renormalization constants into eq.~\eqref{eq:LHSuvr}, its r.h.s. will be expanded and subsequently truncated to $\mathcal{O}(a_s^4)$.

Taking the operator mixing described in eq.~\eqref{eq:FFuvr} into account, the UV renormalization of $\mathcal{M}_{rhs}$ proceeds as follows:
\begin{eqnarray}
\label{eq:RHSuvr}
\mathcal{M}_{rhs} &=& 
Z_{F\tilde{F}} \, Z_3 \, \hat{\mathcal{M}}_{rhs}\left( {\hat a}_s\,,\, \hat{\xi} \right) 
\,+\, Z_{FJ} \, Z_3 \, \hat{\mathcal{M}}_{lhs}\left( {\hat a}_s\,,\, \hat{\xi} \right)\nonumber\\
&=& Z_{F\tilde{F}} \, Z_3 \,  \hat{\mathcal{M}}_{rhs}\left( Z_{a_s}\, a_s\,,\, 1 - Z_3 + Z_3\, \xi  \right) \,+\, Z_{FJ} \, Z_3 \, \hat{\mathcal{M}}_{lhs}\left( Z_{a_s}\, a_s\,,\, 1 - Z_3 + Z_3\, \xi  \right)\nonumber\\
\end{eqnarray}
Comments given above apply here as well.
We note that the renormalization of $\mathcal{M}_{rhs}$ chosen in eq.~\eqref{eq:RHSuvr} is a standard $\MSbar$ renormalization.
For extracting $Z^{f}_{5}$ by matching the two sides of the axial-anomaly equation \eqref{eq:ABJanomalyEQ}, $\mathcal{M}_{rhs}$ is needed to one power less in $a_s$, and hence one-loop less, than that of $\mathcal{M}_{lhs}$, because there is one power of $a_s$ in front of its r.h.s. 
However, besides extracting $Z^{f}_{5}$, we would also like to determine $Z_{F\tilde{F}}$ to $\mathcal{O}(a^4_s)$ by computing $\mathcal{M}_{rhs}$ to 4-loop order. 
To this end, one observes that $Z_{FJ}$ is involved only to $\mathcal{O}(a^3_s)$ since the $\hat{\mathcal{M}}_{lhs}$ has no tree-level diagrams.
Furthermore, since $Z_{FJ}$ itself starts from order $\mathcal{O}(a_s)$, the $\hat{\mathcal{M}}_{lhs}$ is needed only up to 3-loop order in the renormalization of $\hat{\mathcal{M}}_{rhs}$ at the 4-loop order.
Thus, to compute $Z^{f}_{5}$ to $\mathcal{O}(a^3_s)$ and $Z_{F\tilde{F}}$ to $\mathcal{O}(a^4_s)$, all ingredients are known in the literature except for the 4-loop bare expressions of $\mathcal{M}_{lhs}$ and $\mathcal{M}_{rhs}$ which are computed for the first time in this article.

With the analytical expressions of these bare matrix elements to 4-loop order (provided as supplementary material associated with this article) at hand, we present below our results for $Z_J \equiv Z^{ms}_{5} \, Z^{f}_{5}$ and $Z_{F\tilde{F}}$.

\section{Results and discussions}
\label{sec:res}

Whereas the off-shell matrix elements $\mathcal{M}_{lhs}$ and $\mathcal{M}_{rhs}$  depend on the gluon-field gauge-fixing parameter $\xi$, the renormalization constant $Z_J \equiv Z^{ms}_{5} \, Z^{f}_{5}$ of the gauge-invariant axial-current operator $J_5^{\mu}$ is independent of $\xi$. 
As shown in eq.~\eqref{eq:LHSuvr} and eq.~\eqref{eq:RHSuvr}, the $\xi$ dependence of these matrix elements should be kept to 3-loop order\footnote{In fact, to renormalize these off-shell matrix elements at the 4-loop order, their 3-loop expressions are only required to the first power in $\xi$.} in order to renormalize them at 4-loop order.  
We only computed the 4-loop matrix elements in the Feynman gauge $\xi = 0$.

By using eq.~\eqref{eq:LHSuvr}, one can extract the $\MSbar$ renormalization constant $Z^{ms}_{5}$ to $\mathcal{O}(a^3_s)$ from the $\epsilon$ poles of the 4-loop expression of $\hat{\mathcal{M}}_{lhs}$.
We obtain the same result as the one given in refs.~\cite{Larin:1993tq,Ahmed:2015qpa}.
For reader's convenience, we document here the analytic result:
\begin{align}
\label{eq:zms5}
\begin{autobreak}
Z^{ms}_{5} = 
    1      
    + a_s^2 \Big\{ {C_A} {C_F} \Big( \frac{22 }{3 {\epsilon}}\Big) 
    +{C_F} {n_f} \Big( \frac{5 }{3 {\epsilon}} \Big)   \Big\}       
    + a_s^3 \Big\{ C_A^2 {C_F} \Big(\frac{3578}{81 {\epsilon}}-\frac{484}{27 {\epsilon}^2}\Big)    
    + {C_A} {C_F} {n_f} \Big(\frac{149}{81 {\epsilon}}-\frac{22}{27 {\epsilon}^2}\Big)     
    + {C_A} C_F^2 \Big( - \frac{308   }{9 {\epsilon}}\Big)    
    + C_F^2 {n_f} \Big( - \frac{22 }{9 {\epsilon}} \Big)     
    +{C_F} n_f^2 \Big(\frac{20}{27 {\epsilon}^2}+\frac{26}{81 {\epsilon}}\Big)   \Big\} \,.
\end{autobreak}
\end{align}
The definition of the quadratic Casimir color constants is as usual: $C_A = N_c \,, \, C_F = (N_c^2 - 1)/(2 N_c) \,$ along with the color-trace normalization factor $\TR = {1}/{2}$.

In order to determine the non-$\MSbar$ constant $Z^{f}_{5}$ at $\mathcal{O}(a^3_s)$, one needs in addition the $\MSbar$-renormalized $\mathcal{M}_{rhs}$ at 3-loop order inserted into the axial-anomaly equation~\eqref{eq:ABJanomalyEQ}.
The $\MSbar$ renormalized l.h.s. to $\mathcal{O}(a_s^4)$ in Feynman gauge\footnote{$\hat{\mathcal{M}}_{lhs}$ and $\hat{\mathcal{M}}_{rhs}$ with full $\xi$-dependence to 3-loop order can be found in the supplementary material associated with this article.} ($\xi=0$) reads 
\begin{align}
\label{eq:LHSuvr_msbar}
\begin{autobreak}
\bar{\mathcal{M}}_{lhs} = 
 a_s \Big\{4 n_f\Big\} 
    + a_s^2 \Big\{ 24 C_A n_f + 16 C_F n_f\Big\} 
    + a_s^3 \Big\{ C_A^2 {n_f} \Big(16 \zeta_3+\frac{4537}{12}\Big)
    + {C_A} {C_F} {n_f} \Big(\frac{1292}{9}\Big) + C_A n_f^2 \Big(-48 \zeta_3-\frac{343}{6}\Big)-24 C_F^2
   {n_f}+{C_F} n_f^2 \Big(32 \zeta_3-\frac{698}{9}\Big) \Big\}
   + a_s^4 \Big\{ C_A^3 {n_f} \Big(\frac{2822}{9} \zeta_3-500 \zeta_5+\frac{14896805}{1944}+\frac{\pi ^4}{30}\Big)
   +C_A^2 {C_F} {n_f} \Big(\frac{57125}{27}-160 \zeta_3\Big)
   + C_A^2 n_f^2 \Big(-\frac{12752}{9} \zeta_3 +\frac{1600 }{3} \zeta_5 -\frac{1063039}{486}-\frac{\pi ^4}{5}\Big)
   +{C_A} C_F^2 {n_f} \Big(640 \zeta_3-\frac{16952}{27}\Big)
   +{C_A} {C_F} n_f^2 \Big(\frac{4864}{9} \zeta_3+320 \zeta_5-\frac{494545}{162}+\frac{4 \pi ^4}{15}\Big)
   + {C_A} n_f^3 \Big(\frac{368}{3} \zeta_3 +\frac{31021}{243}\Big)
   +C_F^3 {n_f} \Big(\frac{136}{3}-384 \zeta_3\Big)+C_F^2 n_f^2 \Big(\frac{2048 \zeta_3}{3}-640 \zeta_5-\frac{3832}{27}\Big)
   +{C_F} n_f^3 \Big(\frac{19124}{81}-\frac{1024 }{9} \zeta_3\Big)
   \Big\}\,,
\end{autobreak}
\end{align}
and in the same gauge, the $\MSbar$ renormalized r.h.s. is given to $\mathcal{O}(a_s^4)$ by 
\begin{align}
\label{eq:RHSuvr_msbar}
\begin{autobreak}
a_s \, n_f\, \TR \, \mathcal{M}_{rhs} =
a_s \, \frac{n_f}{2} \, 
\Bigg( 
 8
    + a_s \Big\{48 C_A \}    
    + a_s^2 \Big\{ C_A^2 \Big(32 \zeta_3+\frac{4537}{6}\Big)
    +{C_A} {n_f} \Big(-96 \zeta_3-\frac{343}{3}\Big)
    +{C_F} {n_f} \Big(64 \zeta_3-\frac{424}{3}\Big) \Big\}       
    + a_s^3 \Big\{ C_A^3 \Big(\frac{5644}{9} \zeta_3-1000 \zeta_5+\frac{14896805}{972}+\frac{\pi ^4}{15}\Big)   
    +C_A^2 {n_f} \Big(-\frac{25504}{9} \zeta_3+\frac{3200 }{3}\zeta_5-\frac{1063039}{243}-\frac{2 \pi ^4}{5}\Big)   
   +{C_A} {C_F} {n_f} \Big(\frac{15824}{9} \zeta_3+640 \zeta_5-\frac{150623}{27}+\frac{8 \pi ^4}{15}\Big)   
   + {C_A} n_f^2 \Big(\frac{736 }{3}\zeta_3+\frac{62042}{243}\Big)   
   +C_F^2 {n_f} \Big(832 \zeta_3-1280 \zeta_5+\frac{1396}{3}\Big)   
   +{C_F} n_f^2 \Big(\frac{13592}{27}-\frac{2048
   }{9} \zeta_3 \Big) \Big\} \Bigg)\,,
\end{autobreak}
\end{align}
where $\mathcal{M}_{rhs}$ is required only to $\mathcal{O}(a_s^3)$.
Clearly, these two expressions, \eqref{eq:LHSuvr_msbar} and  \eqref{eq:RHSuvr_msbar}, do not agree with each other. 
Without incorporating the finite renormalization constant $Z^{f}_{5}$, the axial-anomaly equation~\eqref{eq:ABJanomalyEQ} is not preserved.
By demanding the validity of eq.~\eqref{eq:ABJanomalyEQ}, the finite renormalization constant $Z^{f}_{5}$ is thus determined to be 
\begin{align}
\label{eq:zf5}
\begin{autobreak}
Z^{f}_{5} =
    1    
    +a_s \Big\{ -4 {C_F} \Big\}   
    + a_s^2 \Big\{ {C_A} {C_F}\Big( -\frac{107 }{9}\Big)+ C_F^2 \Big( 22 \Big) + {C_F} {n_f} \Big( \frac{31 }{18} \Big)  \Big\}       
    + a_s^3 \Big\{ C_A^2 {C_F} \Big(56 \zeta_3-\frac{2147}{27}\Big)    
    + {C_A} C_F^2 \Big(\frac{5834}{27}-160 \zeta_3\Big)    
    + {C_A} {C_F} {n_f} \Big(\frac{110 }{3}\zeta_3-\frac{133}{81}\Big)   
   +C_F^3 \Big(96 \zeta_3-\frac{370}{3}\Big)   
   +C_F^2 {n_f} \Big(\frac{497}{54}-\frac{104 }{3}\zeta_3\Big)   
   + {C_F} n_f^2 \Big( \frac{316 }{81}  \Big) \Big\}\,.
\end{autobreak}
\end{align}
It is obtained by perturbatively expanding the ratio of the \textit{finite} $\bar{\mathcal{M}}_{lhs}$ and $a_s  n_f \TR  \mathcal{M}_{rhs}$ to $\mathcal{O}(a_s^3)$.
The first two orders of eq.~\eqref{eq:zf5} agree with ref.~\cite{Larin:1993tq}, while the third order terms are our new result.
Furthermore, $\bar{\mathcal{M}}_{lhs}$ to 3-loop order and $\mathcal{M}_{rhs}$ to 2-loop order (with full $\xi$ dependence) are checked to be the same as those given in ref.~\cite{Larin:1993tq}, which serves as an indirect confirmation of the statement made in section~\ref{sec:proj}: our projector eq.(\ref{eq:anomalyprojector}) is equivalent to the projection operation devised in ref.~\cite{Larin:1993tq}.
In addition, we have verified that the ratio determined above is independent of $\mu$ by inserting expressions of $\mathcal{M}_{lhs}$ and $\mathcal{M}_{rhs}$ with explicitly restored $\mu$ dependence.
Because $Z_3$ appears as a global factor common on both sides, its incorporation in the calculation is not necessary if one is just interested in determining $Z^{f}_{5}$.
However, without including $Z_3$, the matrix elements of both sides are not fully UV renormalized, and one needs to be careful with truncating these UV-singular expressions to appropriate orders in $\epsilon$. 
The result for $Z^{f}_{5}$ in Quantum Electrodynamics (QED) can be obtained from eq.~\eqref{eq:zf5} by the known substitution rules: $a_s \rightarrow \frac{\alpha}{4 \pi}\,, \frac{n_f}{2} \rightarrow n_f, C_A \rightarrow 0, C_F \rightarrow 1$ where $\alpha$ denotes the QED fine-structure constant and $\frac{n_f}{2} \rightarrow n_f$ is from $\TR \rightarrow 1$.
Finally, we note that the difference between $Z^{f}_{5}$ in eq.~\eqref{eq:zf5} and the additional finite renormalization constant for the non-singlet axial current computed to $\mathcal{O}(a_s^3)$ in ref.~\cite{Larin:1991tj} starts at $\mathcal{O}(a_s^2)$ and is proportional to $n_f\, C_F$.

With both $Z^{f}_{5}$ and $Z^{ms}_{5}$ known to $\mathcal{O}(a^3_s)$, we can now compute the full anomalous dimension $\gJJ$ according to eq.~\eqref{eq:AMDinZs}, 
\begin{align}
\label{eq:AMDofZJ5}
\gJJ &= \epsilon \,+\, \mu^2 \frac{\mathrm{d}\, \ln \left( Z^{f}_{5} \, Z^{ms}_{5}\right)}{\mathrm{d}\, \mu^2} \nonumber\\
     &= \epsilon \,+\, a_s \Big\{ 4 C_F \epsilon \Big\} 
    + a_s^2 \Big\{  {C_A} {C_F} \Big( \frac{214  }{9} {\epsilon} \Big)- C_F^2 \Big( 28  {\epsilon}\Big) +{C_F} {n_f} \Big(-\frac{31 }{9}{\epsilon}-6\Big)  \Big\} \nonumber\\
    &+ a_s^3 \Big\{ C_A^2 {C_F} \Big(\frac{2147 }{9}{\epsilon}-168 {\epsilon} \zeta_3\Big)
    + {C_A} C_F^2 \Big(480 {\epsilon} \zeta_3-\frac{4550 }{9} {\epsilon}\Big) \nonumber\\
    &+ {C_A} {C_F} {n_f}    \Big(-110 {\epsilon} \zeta_3+\frac{133 }{27}{\epsilon}-\frac{142}{3}\Big)
   +C_F^3 (170 {\epsilon}-288 {\epsilon} \zeta_3) \nonumber\\
   &+C_F^2 {n_f} \Big(104 {\epsilon} \zeta_3-\frac{869 }{18} {\epsilon}+18\Big)
   +{C_F} n_f^2 \Big(\frac{4}{3}-\frac{316}{27}  {\epsilon}\Big)  \Big\}\,.
\end{align}
The 4-dimensional limit of $\gJJ$ was determined in ref.~\cite{Larin:1993tq}, with which we agree.
The $\epsilon$-dependent parts in eq.~\eqref{eq:AMDofZJ5} to $\mathcal{O}(a_s^2)$ agree with those given in ref.~\cite{Ahmed:2015qpa}, apart from a pure $\epsilon$ term at $\mathcal{O}(a_s^0)$ due to the $\mu^{2\epsilon}$ factor introduced in eq.~(\ref{eq:Zsmatrix}).
We stress that the full form of $\gJJ$ given by eq.~\eqref{eq:AMDofZJ5}, including its $\epsilon$-dependent terms, should be used when trying to solve eq.~\eqref{eq:AMDinZs} for renormalization constants.
As far as $\gJJ$ at $\mathcal{O}(a^3_s)$ is concerned, the newly found $\mathcal{O}(a^3_s)$ terms of $Z^{f}_{5}$ contribute only to its $\epsilon$-dependent part. 
However, when $\gJJ$ will be computed at $\mathcal{O}(a^4_s)$ these terms will affect its 4-dimensional part as well.  
~\\

The matrix element $\mathcal{M}_{rhs}$ is only needed to 3-loop order in eq.~\eqref{eq:RHSuvr_msbar} for the computation of $Z^{f}_{5}$.
With its 4-loop expression at hand, we have determined the $Z_{F\tilde{F}}$ constant to $\mathcal{O}(a^4_s)$ according to eq.~\eqref{eq:RHSuvr}, where this renormalization constant is the only one involved to $\mathcal{O}(a^4_s)$ as explained before.
Subsequently, we have verified explicitly that 
\begin{eqnarray}
\label{eq:ZFFidentity}
Z_{F\tilde{F}} = Z_{a_s}  
\end{eqnarray}
holds true at the 4-loop order (i.e.~$\mathcal{O}(a^4_s)$).
This equality was shown in ref.~\cite{Breitenlohner:1983pi} to ensure the extension of the Adler-Bardeen theorem to a non-abelian gauge theory. 
The equality was further claimed to hold true in the same reference by showing that, apart from the mixing with the divergence of the axial-current operator, the operator $\hat{a}_s \, \big[F\tilde{F}\big]$ does not need any multiplicative renormalization in a gauge-invariant regularization scheme using the background field method~\cite{Abbott:1980hw}.
We refrain from reproducing here the 4-loop expression of $Z_{F\tilde{F}}$, which contains three non-quadratic color constants, as it is checked to be in perfect agreement with $Z_{a_s}$ (whose expression can be found in refs.~\cite{vanRitbergen:1997va,Chetyrkin:1999pq,Czakon:2004bu,Chetyrkin:2004mf}). 
Curious readers are invited to repeat this check in accordance with eq.~\eqref{eq:RHSuvr} using the bare matrix elements provided in the accompanying supplemental material.
We note that among the three non-quadratic color constants involved in this calculation, the color factor\footnote{
The symmetric tensor $d_F^{abcd}$ is defined by the color trace $\frac{1}{6} \mbox{Tr}$$\big( T^a T^b T^c T^d + T^a T^b T^d T^c + T^a T^c T^b T^d + T^a T^c T^d T^b + T^a T^d T^b T^c + T^a T^d T^c T^b \big)$ with $T^a$ the generators of the fundamental representation of the SU($N_c$) group.} $d_F^{abcd} d_F^{abcd} = (N_c^2-1)(N_c^4 - 6 N_c^2 + 18)/(96 N_c^2)$ is absent in the $\frac{1}{\epsilon}$ pole of the bare 4-loop $\mathcal{M}_{rhs}$. 
It appears, however, both in $Z_{F\tilde{F}}$ and $Z_3$, but cancels in the product $Z_{F\tilde{F}} \, Z_3$.
The Abelian version of eq.~\eqref{eq:ZFFidentity} was known already since the early reference~\cite{Adler:1969gk}.
This relation was first verified explicitly in QCD to $\mathcal{O}(a_s)$ in ref.~\cite{Espriu:1982bw}, to $\mathcal{O}(a^2_s)$ in ref.~\cite{Bos:1992nd,Larin:1993tq} and subsequently to $\mathcal{O}(a^3_s)$ in refs.~\cite{Zoller:2013ixa,Ahmed:2015qpa}, and now finally in this article to $\mathcal{O}(a^4_s)$ where non-quadratic color Casimirs start to appear. 

Before we move on, it is interesting to note that with eq.~\eqref{eq:ZFFidentity} the renormalization constant in front of $\hat{\mathcal{M}}_{rhs}$ in eq.~\eqref{eq:RHSuvr} reduces to unity in QED owing to the special all-order equality between the fermion field renormalization constant $Z_2$ and the QED-vertex renormalization constant $Z_1$. 
To be more specific, in QED it reads $Z_{F\tilde{F}} \, Z_3 = Z_{\alpha} \, Z_3 = \left(\frac{Z_1}{Z_2} \right)^2 = 1$. 
In consequence, the only UV counterterms needed in this case to renormalize $\hat{\mathcal{M}}_{rhs}$  are due to the operator mixing of $F\tilde{F}$ with $\partial_{\mu}J^{\mu}_5$.    
~\\

Subjecting both sides of the axial-anomaly equation~\eqref{eq:ABJanomalyEQ} to the logarithmic derivative $\mu^2 \frac{\mathrm{d}\, }{\mathrm{d}\, \mu^2}$, one obtains~\cite{Larin:1993tq}  
\begin{eqnarray}
\label{eq:ABJanomalyEQsAMD}
\left(\gJJ - n_f \, \TR \, a_s\, \gFJ - \gFF - \beta + \epsilon \right) \big[\partial_{\mu} J^{\mu}_{5}\big]_{R} = 0\,,
\end{eqnarray}
with anomalous dimensions defined in eq.~\eqref{eq:AMDinZs}, which actually holds only in the 4-dimensional limit $\epsilon = 0$.
The equality between the $\MSbar$ renormalization constants, $Z_{F\tilde{F}}$ and $Z_{a_s}$, explicitly verified to 4-loop order, implies the equality $\gFF = -\beta + \epsilon = -\mu^2\frac{\mathrm{d} \ln a_s}{\mathrm{d} \mu^2}$, to the same perturbative order.
Clearly, eq.~\eqref{eq:ABJanomalyEQsAMD}, which is a natural consequence of the all-order axial-anomaly equation~\eqref{eq:ABJanomalyEQ}, combined with the equality $\gFF = -\beta + \epsilon$ necessarily implies that 
\begin{eqnarray}
\label{eq:AMDsEQ}
\gJJ = n_f \, \TR \, a_s\, \gFJ 
\end{eqnarray}
must hold true, albeit only in the limit $\epsilon = 0$.
This relation has been checked by explicit calculations to $\mathcal{O}(a^2_s)$ in ref.~\cite{Larin:1993tq} and  $\mathcal{O}(a^3_s)$ in refs.~\cite{Zoller:2013ixa,Ahmed:2015qpa}, and discussed in refs.~\cite{Matiounine:1998re,Vogt:2008yw,Moch:2014sna,Behring:2019tus}.
This relation was also shown in ref.~\cite{Breitenlohner:1983pi,Bos:1992nd} to be natural in a non-Abelian extension of the Adler-Bardeen theorem.

Reformulating the axial-anomaly equation~\eqref{eq:ABJanomalyEQ} in terms of the bare fields, one arrives at 
\begin{eqnarray}
\label{eq:ABJanomalyEQbare}
\mu^{2\epsilon} \big(\, Z_{J} - n_f \, \TR \, a_s\, Z_{FJ} \big) \big[\partial_{\mu} J^{\mu}_{5}\big]_{B} = \hat{a}_s \, n_f \, \TR\, \big[F \tilde{F}\big]_{B}
\end{eqnarray}
where we note the appearance of the factor $\mu^{2\epsilon}$, as well as the RG-invariant bare coupling $\hat{a}_s$, and all terms have their canonical mass-dimensions as in four dimensions.
Applying the logarithmic derivative in $\mu^2$ to the axial current on the l.h.s. of eq.~(\ref{eq:ABJanomalyEQbare}), we obtain 
\begin{eqnarray}
\label{eq:RGIaxialcurrentAMD}
\mu^2\frac{\mathrm{d}}{\mathrm{d} \mu^2} \Big(\mu^{2\epsilon} \big( Z_{J} - n_f \, \TR \, a_s\, Z_{FJ} \big) \Big) \big[J^{\mu}_{5}\big]_{B}
= \big(\gJJ \,-\, n_f \, \TR \, a_s\, \gFJ \big) \big[J^{\mu}_{5}\big]_{R}\,,
\end{eqnarray}
which is explicitly satisfied with the anomalous dimensions determined in the present publication but only in the 4-dimensional limit owing to eq.~(\ref{eq:AMDsEQ}).
Clearly, the RG-invariance of the operator on the r.h.s. of eq.~(\ref{eq:ABJanomalyEQbare}) is exact in $D$ dimensions as long as the Levi-Civita tensor in the definition of $F\tilde{F}$ is algebraically consistently defined. The RG-invariance of the l.h.s.~operator, on the other hand, is only conditionally true in the 4-dimensional limit. 
This reflects the fact that eq.~\eqref{eq:ABJanomalyEQ}, as well as eq.~\eqref{eq:ABJanomalyEQbare}, should be regarded as a relation valid in the 4-dimensional limit, rather than a $D$-dimensional identity.
It was shown by explicit calculations in ref.~\cite{Espriu:1982bw} that the coefficient $\big(\, Z_{J} - n_f \, \TR \, a_s\, Z_{FJ} \big)$ in front of the bare axial current $\big[J^{\mu}_{5}\big]_{B}$ on the l.h.s of eq.~\eqref{eq:ABJanomalyEQbare} reduces to unity up to $\mathcal{O}(a_s^2)$ in QCD with the $\gamma_5$ prescription of ref~\cite{Chanowitz:1979zu}.
Apparently, with Larin's prescription of $\gamma_5$~\cite{Larin:1991tj,Larin:1993tq}, this coefficient is no longer unity, but the axial current in eq.~(\ref{eq:ABJanomalyEQbare}) remains RG-invariant, albeit only in the 4-dimensional limit.

\section{Conclusion}
\label{sec:conc}

We have extended the knowledge of the renormalization constant $Z_J$, and in particular, of the finite non-$\MSbar$ factor $Z_5^f$, of the flavor-singlet axial-vector current regularized with Larin's prescription of $\gamma_5$ to $\mathcal{O}(\alpha_s^3)$ in QCD through computations of matrix elements of operators appearing in the axial-anomaly equation $\big[\partial_{\mu} J^{\mu}_{5} \big]_{R} = a_s\, n_f\, \TR \,  \big[F \tilde{F} \big]_{R}$ between the vacuum and a pair of (off-shell) gluons to 4-loop order.  
Furthermore, we have verified explicitly up to 4-loop order the equality between the $\MSbar$ renormalization constant $Z_{F\tilde{F}}$ associated with the operator $\big[F \tilde{F} \big]_{R}$ and that of $\alpha_s$.
This equality automatically ensures that the relation $\gJJ = a_s\, n_f\, \TR \, \gFJ $ remains valid to $\mathcal{O}(\alpha_s^4)$ given the correctness of the all-order axial-anomaly equation which was used to determine the non-$\MSbar$ piece of $Z_J$ in Larin's prescription of $\gamma_5$.

~\\
\textit{Note added}: While this work was under review, a proof of the absence of a (divergent) multiplicative renormalization of the axial-anomaly operator in dimensionally regularized QCD was recently completed in ref.~\cite{Luscher:2021bog} in Becchi-Rouet-Stora quantization~\cite{Becchi:1974md,Becchi:1975nq}. In a private communication, the authors of ref.~\cite{Luscher:2021bog} pointed to us that the proof in ref.~\cite{Breitenlohner:1983pi} is incomplete.

\section*{Acknowledgements}

L.C. would like to thank M.~Niggetiedt for a helpful chat regarding the organization of the diagram generation.
The work of L.C. and M.C. was supported by the Deutsche Forschungsgemeinschaft under grant 396021762 -- TRR 257.
The work of T.A. received funding from the European Union’s Horizon 2020 research and innovation programme under grant agreement No 772099.
We gratefully acknowledge the computing resources provided by Max-Planck-Institute for Physics and by RWTH Aachen University where some of the computations have been completed.

\appendix
\section{Derivation of the axial-anomaly projector}
\label{append:aap}

Here, we give some details on the conditional equivalence between the projector $\mathcal{P}_{\mu \mu_1 \mu_2}$ in eq.~(\ref{eq:anomalyprojector}), which we use, and the projection operation devised for projecting out the anomaly in eq.~(19) of ref.~\cite{Larin:1993tq} as far as the rank-3 matrix element $\Gamma^{\mu \mu_1 \mu_2}_{lhs}(p_1, p_2)$ defined in eq.~(\ref{eq:Glhs1PI}) is concerned.
First of all, let us recap the r.h.s. of eq.~(19) of ref.~\cite{Larin:1993tq}, which reads
\begin{eqnarray}
\label{eq:LsP0}
\mathrm{M} = 
R_{\MSbar} \, \epsilon_{\mu_1 \mu_2 \rho \sigma} \, \frac{p_1^{\rho}}{p_1^2}\, \frac{\partial }{\partial q_{\sigma}} \, \int d^4 x  d^4 y \, e^{-i p_1 \cdot x - i q \cdot y }\,  \langle 0|\hat{\mathrm{T}}\left[ \frac{\partial }{\partial y^{\mu}} J_{5}^{\mu}(y) \, A_a^{\mu_1}(x) \, A_a^{\mu_2}(0) \right] |0 \rangle |_{\mathrm{amp}} \,,\nonumber\\
\end{eqnarray}
where the notations and conventions have been matched to those adopted in this article, and $R_{\MSbar}$ denotes the R-operation in the $\MSbar$ scheme. 
Implicit in the following manipulations of the expression in eq.~(\ref{eq:LsP0}) is that all possible divergences therein are properly regularized, and no additional ``ill-defined'' divergence is generated at intermediate steps.
This is ensured by the usage of dimensional regularization, $\MSbar$ subtraction and also by the fact that there is no IR divergence in the three-point correlation function involved at the chosen kinematic point (illustrated in figure~\ref{fig:kinematics}) due to the IR-rearrangement by setting both gluons off-shell.

Since the equal-time commutator between $J_{5}^{0}(y) = \sum_{\psi} \bar{\psi}(y) \gamma^0 \gamma_5 \psi(y)$ and $A_a^{\mu}(x)$ vanishes, the derivative w.r.t $y$ can be pulled in front of the integral, and we arrive at  
\begin{eqnarray}
\label{eq:LsP1}
\mathrm{M} &=&
R_{\MSbar} \, \epsilon_{\mu_1 \mu_2 \rho \sigma} \, \frac{p_1^{\rho}}{p_1^2}\, \frac{\partial }{\partial q_{\sigma}} \, \int d^4 x  d^4 y \, e^{-i p_1 \cdot x - i q \cdot y }\, \frac{\partial }{\partial y^{\mu}}  \langle 0|\hat{\mathrm{T}}\left[ J_{5}^{\mu}(y) \, A_a^{\mu_1}(x) \, A_a^{\mu_2}(0) \right] |0 \rangle |_{\mathrm{amp}} \nonumber\\
&=&
R_{\MSbar} \, \epsilon_{\mu_1 \mu_2 \rho \sigma} \, \frac{p_1^{\rho}}{p_1^2}\, \frac{\partial }{\partial q_{\sigma}} \, \int d^4 x  d^4 y \, e^{-i p_1 \cdot x - i q \cdot y }\, \nonumber\\
&& \frac{\partial }{\partial y^{\mu}} 
\left( \int \frac{d^4 k_1}{(2 \pi)^4} \frac{d^4 Q}{(2 \pi)^4} 
\, e^{i k_1 \cdot x + i Q \cdot y }\, 
\Gamma^{\mu \mu_1 \mu_2}_{lhs}(k_1, -Q-k_1)  
\right) \nonumber\\
&=&
R_{\MSbar} \, \epsilon_{\mu_1 \mu_2 \rho \sigma} \, \frac{p_1^{\rho}}{p_1^2}\,\frac{\partial }{\partial q_{\sigma}} \,
\Big( 
i q_{\mu} \Gamma^{\mu \mu_1 \mu_2}_{lhs}(p_1, q-p_1)  
\Big)\nonumber\\
&=&
R_{\MSbar} \, \epsilon_{\mu_1 \mu_2 \rho \sigma} \, \frac{ p_1^{\rho}}{p_1^2}\,
i \Big( 
\Gamma^{\rho \mu_1 \mu_2}_{lhs}(p_1, q-p_1)  
+ 
q_{\mu} \frac{\partial }{\partial q_{\sigma}} \Gamma^{\mu \mu_1 \mu_2}_{lhs}(p_1, q-p_1)
\Big)
\,,
\end{eqnarray}
where in the second line we have inserted the Fourier transformation of the correlation function in coordinate space in terms of the momentum-space matrix element $\Gamma^{\mu \mu_1 \mu_2}_{lhs}(p_1, p_2)$ defined in eq.~(\ref{eq:Glhs1PI}).
The first term in the last line of eq.~\eqref{eq:LsP1} corresponds to the projection we are using in eq.~(\ref{eq:anomalyprojector}), up to an overall factor. 
The second term, proportional to $\epsilon_{\mu_1 \mu_2 \rho \sigma} \, p_1^{\rho}\,q_{\mu}\, \frac{\partial }{\partial q_{\sigma}}\, \Gamma^{\mu \mu_1 \mu_2}_{lhs}(p_1, q-p_1)$, should vanish in the limit $q \rightarrow 0$, as long as the derivative $\frac{\partial }{\partial q_{\sigma}} \Gamma^{\rho \mu_1 \mu_2}_{lhs}(p_1, q-p_1)$ has no power divergence in this limit (i.e.~is not more singular than $|q|^{-1}$ as $q \rightarrow 0$).

With all UV divergences in $\Gamma^{\mu \mu_1 \mu_2}_{lhs}(p_1, p_2)$ regularized dimensionally, which can all be factorized and subtracted by momentum/mass-independent $\MSbar$ renormalization constants, the coefficients of its Laurent expansion series in $\epsilon$ are expected to behave regularly, i.e.~not be divergent, when $q \rightarrow 0$ with $p_1^2$ fixed at an off-shell point (as no IR divergence is present here). 
Under this regularity condition on $\Gamma^{\mu \mu_1 \mu_2}_{lhs}(p_1, p_2)$ regarding its asymptotic behavior at $q \rightarrow 0$, we see that the projector $\mathcal{P}_{\mu \mu_1 \mu_2}$ in eq.~(\ref{eq:anomalyprojector}) should project out the same object as the operation in eq.~(\ref{eq:LsP0}).
The fact that we obtained the same $\bar{\mathcal{M}}_{lhs}$ to 3-loop order and $\mathcal{M}_{rhs}$ to 2-loop order (with full $\xi$ dependence) as those given in ref.~\cite{Larin:1993tq}, as discussed in section~\ref{sec:res}, indirectly confirms this point.

\bibliography{Z5singlet} 
\bibliographystyle{utphysM}
\end{document}